\documentclass [12pt]{article}
\usepackage [cp1251]{inputenc}
\usepackage [english]{babel}
\usepackage {graphicx}
\usepackage {amssymb}
\usepackage {amsmath}
\usepackage {longtable}
\title{NONLINEAR VORTEX STRUCTURES IN OBLIQUELY ROTATING STRATIFIED
 FLUIDS   DRIVEN BY SMALL SCALE NON HELICAL FORCES}
\author{$^{1}$\textbf{M.I. Kopp}, $^3$\textbf{A.V. Tur}, $^{1,2}$\textbf{V.V. Yanovsky}}

\begin{document}

 \maketitle

$^{1}$ \textit{Institute for Single Crystals, NAS  Ukraine, Nauky Ave. 60, Kharkov 61001, Ukraine }

$^{2}$\textit{ V.N. Karazin Kharkiv National University 4 Svobody Sq., Kharkov 61022, Ukraine}

$^{3}$\textit{Universit\'{e} de Toulouse [UPS], CNRS, Institut de Recherche en Astrophysique \\et Plan\'{e}tologie,
9 avenue du Colonel Roche, BP 44346, 31028 Toulouse Cedex 4, France}

\bigskip

\begin{abstract}In this paper, we study a new type of large-scale instability in obliquely rotating stratified fluids with small scale non-helical turbulence. The small-scale turbulence is generated by the external force with zero helicity and low Reynolds number. The theory uses the method of multiscale asymptotic developments. The nonlinear equations for large scale motions are obtained in the third order of the perturbation theory. In this paper, we consider the linear instability and the stationary nonlinear modes. We obtain solutions in the form of nonlinear Beltrami waves and localized vortex structures as kinks of new type.
\end{abstract}

\section{Introduction}
The problem of generation of large-scale vortex motions and formation of stationary nonlinear vortex structures in turbulent media is very important.  The characteristic scales of the velocity fields for these structures are much bigger than the characteristic scales of turbulent motions or waves which engender their appearance. The study of the generation mechanism of large-scale vortex structures (LSVS) by small-scale turbulence has not only scientific but also application interest. For instance, tornados, cyclones and anticyclones represent the particular cases of LSVS. They also play an important role in the dynamics of Earth's atmosphere, since they determine the global transport of air masses and are responsible for the weather and climate on our planet \cite{1s}-\cite{3s}. The study of the mechanisms of  LSVS generation is important for a number of astrophysical problems,  such as the origin of the Great Spot of Jupiter, Venus super rotation, vortex structures in solar prominences and so on \cite{4s}-\cite{8s}. The generation of LSVS in atmospheres and bowels of the planets (or other space objects) is essentially due to thermal phenomena occurring under the influence of internal or external sources of the thermal energy. Articles \cite{9s}-\cite{11s} present the theory of convective vortex dynamo. It follows from this theory that the small-scale helical turbulence leads to a large-scale instability, which engenders the formation of one convective cell   considered as a huge vortex of type of tropical cyclone. Number of numerical \cite{12s} and laboratory \cite{13s} experiments confirm this.  The theory of the convective vortex dynamo was further developed in \cite{14s}, \cite{15s}, applying the method of multiscale asymptotic developments. This method, unlike the functional techniques \cite{16s}, \cite{17s} used in \cite{9s}-\cite{11s} allows to determine strictly the principal order of perturbations theory in which the instability arises. The method of multiscale asymptotic developments to describe the generation of LSVS in reflective non invariant turbulence was first used in \cite{18s}. The small Reynolds number is a parameter of the asymptotic developments. Nonlinear stabilization of the convective large-scale instability discussed in \cite{14s},\cite{15s} leads to the formation of helical vortex solitons or kinks of a new type. The generation of helical turbulence in natural conditions is usually associated with the influence of the Coriolis force on turbulent motion \cite{19s}, \cite{20s}. Obviously, the question arises of the possibility of generation of large-scale vortex field in a rotating media under influence of small-scale force with zero helicity. An example of the generation of LSVS in a rotating incompressible fluid was found in \cite{21s}. It was also shown that the development of large-scale instability in an inclined rotating fluid generates the nonlinear large-scale helical vortex structures or localized Beltrami vortices with the internal helical structure.

In this paper we give a generalization of the $\alpha $ -- effect found in \cite{21s} for the case of temperature- stratified fluid. As a result of this generalization, we obtain the large-scale instability which generates the LSVS.

The organization of this paper is as follows. In the Section 2 we obtain the averaged hydrodynamic equations in the Boussinesq approximation in an obliquely rotating fluid for the large-scale fields using the method of multiscale asymptotic developments. The technical aspect of this question is described in detail in Appendix I. The correlation functions in the averaged equations are expressed in terms of the small-scale fields in the zero-order approximation with respect to $R$. In Appendix II in order to obtain the averaged equations in closed form, we find solutions for small-scale fields of zero order approximation. In Appendix III we calculate the Reynolds stresses using these solutions. In Sec. 2 we obtain a closed system of the nonlinear equations for large-scale vortex fields (vortex dynamo). In Section 3 we consider the generation of small large-scale vortex perturbations, which arises from instability of the type $\alpha$ -- effect. Also the conditions of the appearance of this instability are determined depending on the effects of rotation and stratification of the medium. In Section 4, an analytic and numerical analysis of nonlinear equations in the steady state is performed, which shows the existence of nonlinear Beltrami waves and localized vortex structures in the form of kinks. The results obtained in the work can be applied to the numerous geophysical and astrophysical problems.

\section{  Equations for the large-scale fields}

Let us consider a system of equations for the perturbations of velocity $\vec{v}$, temperature  $T$, pressure $P$ in the Boussinesq approximation  with the constant temperature gradient $\nabla \overline{T}$  for a rotating coordinate system:
\begin{equation} \label{eq1}
  \frac{\partial v_{i} }{\partial t} +v_{k} \frac{
\partial v_{i} }{\partial x_{k} } =\nu \Delta v_{i} -\frac{1}{\overline{\rho }} \frac{
\partial P}{\partial x_{i} } +2\varepsilon _{ijk} v_{j} { \Omega }_{k} +ge_{i}
\beta T+F_{0}^{i}
\end{equation}

\begin{equation} \label{eq2}
  \frac{\partial T}{\partial t} +v_{k} \frac{
\partial T}{\partial x_{k} } -Ae_{k} v_{k} =\chi \Delta T
\end{equation}
\begin{equation} \label{eq3}
  \frac{\partial v_{i} }{\partial x_{i} } =0
\end{equation}

The system of equations (\ref{eq1})-(\ref{eq3}) describes the evolution of perturbations on the background of the basic equilibrium state $\overline{T}\left(z\right)$,  $\overline{\rho}\left(z\right)$, which is set by the constant temperature gradient $\nabla \overline{T}=-A\vec{e}(A>0)$ (heated from below) and by the hydrostatic pressure: $\nabla \overline{P}=\overline{\rho }\vec{g}-\overline{\rho }\left[\vec{\Omega }\times \left[\vec{\Omega }\times \vec{r}\right]\right]$, where $\vec{r}$ is the radius-vector of the fluid element. We consider the angular velocity of rotation $\vec{\Omega }$ as constant (solid rotation) and inclined to the plane $(X,Y)$, as shown in Fig. 1, i.e. for the Cartesian geometry problem: $\vec{\Omega }=\left(\Omega _{1} ,\Omega _{2} ,\Omega _{3} \right)$ . Where $\vec{e}=(0,0,1)$ is  unit vector in the direction of the axis $Z$, $\vec{g}$ is the gravity directed vertically downwards $\vec{g}=(0,0,-g)$, $\beta $ is the coefficient of thermal expansion. The equation (\ref{eq1}) includes the external force $\vec{F}_{0} $, that models the excitation source in the environment of small scale and high frequency fluctuations of the velocity field with small Reynolds number. Unlike the  previous studies \cite{14s}, \cite{15s} we consider here the non-helical external force with the following properties:

\begin{equation} \label{eq4}
div\vec{F}_{0} =0, \; \vec{F}_{0} rot\vec{F}_{0}
=0,\;  rot\vec{F}_{0} \ne 0 , \;    \vec{F}_{0} =f_{0} \vec{F}_{0} \left(\frac{x}{\lambda
_{0} } ;  \frac{t}{t_{0} } \right)
\end{equation}

Where $\lambda _{0} $ is the characteristic scale, $t_{0}$ is the characteristic time, $f_{0} $ is the characteristic amplitude. The external force is given in the plane $(X,Y)$ , where the perpendicular projection of the angular velocity is located. We choose the external force in rotating coordinate system in the following form :

\begin{equation} \label{eq5}
F_{0} ^{z} =0,\;\vec{F}_{0\bot } =f_{0} \left(\vec{i}cos\phi _{2} +\vec{j}cos\phi _{1}
\right),\; \phi _{1} =\vec{k}_{1} \vec{x}-\omega _{0} t,\phi _{2}
=\vec{k}_{2} \vec{x}-\omega _{0} t,$$
$$\vec{k}_{1}=k_{0} \left(1,0,0\right),\vec{k}_{2} =k_{0} \left(0,1,0\right).
\end{equation}

\begin{figure}
  \centering
    \includegraphics[width=5 cm]{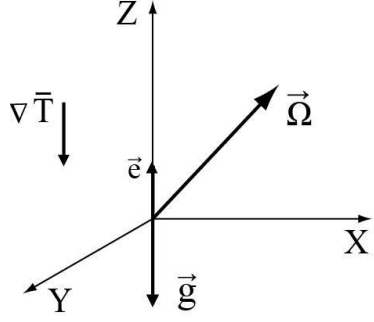}\\
  \caption{ In general case, the angular velocity $\vec{\Omega} $ is inclined to the plane $(X, Y)$ where the external force  $\vec{F}_{0\bot} $ is located.}\label{fg1}
\end{figure}

Obviously, this non-helical external force satisfies all the conditions (\ref{eq4}). Let us now use in equations (\ref{eq1})-(\ref{eq3}) the dimensionless variables.  For convenience, we will designate these variables like dimensional variables:
\[\vec{x}\to \frac{\vec{x}}{\lambda _{0} } , \quad t
\to \frac{t}{t_{0} } , \quad \vec{v}\to \frac{\vec{v}}{v_{0} }  , \quad \vec{F}_{0} \to \frac{
\vec{F}_{0} }{f_{0} } , \quad P\to \frac{P}{\overline{\rho }p_{0} } , \]

\begin{equation} \label{eq6}
  t_{0} =\frac{
\lambda _{0}^{2} }{\nu } , \quad p_{0} =\frac{v_{0} \nu }{\lambda _{0} } , \quad f_{0} =\frac{v_{0}
\nu }{\lambda _{0}^{2} } , \quad T\to \frac{T}{\lambda _{0} A}
\end{equation}

Then, in dimensionless variables the equation (\ref{eq1})-(\ref{eq3}) takes the following form:

\begin{equation} \label{eq7}
  \frac{\partial v_{i} }{\partial t} +Rv_{k} \frac{
\partial v_{i} }{\partial x_{k} } =\Delta v_{i} -\frac{\partial P}{\partial x_{i}
} +\varepsilon _{ijk} v_{j} D_{k} +e_{i}\widetilde{Ra}T+F_{0}^{i}
\end{equation}

\begin{equation} \label{eq8}
  \frac{\partial T}{\partial t} +Rv_{k} \frac{\partial T}{\partial
x_{k} } -e_{k} v_{k} =Pr^{-1} \Delta T
\end{equation}

\begin{equation} \label{eq9}
  \frac{\partial v_{i} }{\partial x_{i} } =0
\end{equation}

Here, $\widetilde{Ra}=\frac{Ra}{Pr} $, $Ra=\frac{g\beta A\lambda _{0}^{4} }{\nu \chi } $ is the Rayleigh number for the scale $\lambda _{0} $,  $Pr=\frac{\nu }{\chi } $  is the  Prandtl number, $D_{i} =\frac{2\Omega _{i} \lambda _{0}^{2} }{\nu } $ is the dimensionless parameter of rotation for the scale $\lambda _{0} $ related with the Taylor number  $Ta_{i} =D_{i}^{2} $ which is the characteristic  of  the influence of  Coriolis force on viscous forces.

We will consider the Reynolds number $R=\frac{v_{0} t_{0} }{\lambda _{0} }  \ll 1$  on the scale $\lambda _{0} $ as a small parameter of an asymptotic development. Concerning the parameters $D_{i} $ and  $Ra$ , we do not choose any range of values for the moment. Let us examine the following formulation of the problem. We consider the external force as small and of high frequency. This force leads to the small scale fluctuations in velocity and temperature fields on the equilibrium background. After averaging, these quickly oscillating fluctuations vanish. Nevertheless, due to the small nonlinear interactions in some orders of perturbation theory, nonzero terms can occur after averaging. This means that they are not oscillatory and are of large scale. From a formal point of view, these terms are secular, i.e., they create the conditions for the solvability of large scale asymptotic development. So the purpose of this paper is to find and study the solvability equations, i.e., the equations for large scale perturbations. The method of asymptotic equations is well presented in works \cite{7s}, \cite{14s}, \cite{15s}. In accordance with these papers we introduce spatial and temporal derivatives in equations (\ref{eq7})-(\ref{eq9}) in the form of asymptotic expansions:
\begin{equation} \label{eq10}
 \frac{\partial }{\partial t} \to \partial _{t}
+R^{4} \partial _{T}, \;\;\; \frac{\partial }{\partial x_{i} } \to \partial _{i} +R^{2}
\nabla _{i}
\end{equation}

where$\partial _{i} $ and  $\partial _{t} $ denote derivatives with respect to fast variables $x_{0} =\left(\vec{x}_{0} ,t_{0} \right)$  and $\nabla _{i} $,$\partial _{T} $ derivatives    with respect to slow variable $X=\left(\vec{X},\; T\right)$ . Variables $x_{0} $ and   $X$ can be called small-scale and large-scale variables. To construct the nonlinear theory, the variables $\vec{v}$, $\vec{T}$, $P$  are presented in the form of asymptotic series:
\[\vec{v}(\vec{x},t) =\frac{1}{R} \vec{W}_{-1} \left(X\right)+\vec{v}_{0} \left(x_{0} \right)+R\vec{v}_{1} +R^{2} \vec{v}_{2} +R^{3} \vec{v}_{3}+\cdots  \]
\begin{equation} \label{eq11}
T(\vec{x},t)=\frac{1}{R} T_{-1} \left(X\right)+T_{0} \left(x_{0} \right)+RT_{1} +R^{2} T_{2} +R^{3} T_{3} +\cdots
\end{equation}
\[P(\vec{x},t)=\frac{1}{R^{3} } P_{-3} +\frac{1}{R^{2} } P_{-2} +\frac{1}{R} P_{-1} +P_{0} +R(P_{1} +\overline{P}_{1} \left(X\right))+R^{2} P_{2} +R^{3} P_{3} +\cdots\]

Substituting developments (\ref{eq10})-(\ref{eq11}) into the initial equations (\ref{eq7})-(\ref{eq9}) and then gathering together the terms of the same order, we obtain the equations of the multi-scale asymptotic development and write down the obtained equations up to order $R^{3}$ including.  The algebraic structure of the asymptotic development of equations (\ref{eq7})-(\ref{eq9}) in various orders of $R$ is given in Appendix I. It is also shown that in order  $R^{3} $   we get the main secular equation or equation for the large-scale fields:
\begin{equation} \label{eq12} \partial _{T} W_{-1}^{i} -\Delta W_{-1}^{i}
+\nabla _{k} \left(\overline{v_{0}^{k} v_{0}^{i} }\right)=-\nabla _{i} \overline{P}_{1}  \end{equation}

\begin{equation}
\label{eq13} \partial _{T} T_{-1} -Pr^{-1} \Delta T_{-1} =-\nabla _{k} \left(
\overline{v_{0}^{k} T_{0} }\right) \end{equation}
Equations (\ref{eq12})-(\ref{eq13}) with secular equations are obtained in Appendix I:
\[ -\nabla _{i} P_{-3} +\widetilde{Ra}e_{i} T_{-1}
+\varepsilon _{ijk} W_{j} D_{k} =0 \]

\begin{figure}
  \centering
    \includegraphics[width=5 cm]{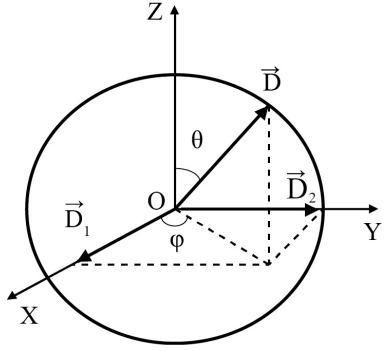}\\
  \caption{Relation of Cartesian projections of rotation parameter  $\vec{D}$ (or angular velocity  of rotation  $\vec{\Omega}$) with their projections in a spherical coordinate system.}\label{fg2}
\end{figure}

\[ W_{-1}^{k} \nabla _{k} W_{-1}^{i} =-\nabla
_{i} P_{-1}  \]

\[ W_{-1}^{k} \nabla _{k} T_{-1} =0 \]

\[ \nabla _{i} W_{-1}^{i} =0   \]

\[ \ W_{-1}^{z}=0  \]
These equations are satisfied by choosing the following geometry for the velocity field:
\begin{equation} \label{eq14}
 \vec{W}_{-1} =\left(W_{-1}^{x} \left(Z\right), \quad W_{-1}^{y}
\left(Z\right),0\right),  \quad T_{-1} =T_{-1} \left(Z\right), \quad P_{-1}
=const
\end{equation}
In the frame of this quasi-two-dimensional approximation, we assume that the large-scale derivative over $Z$ is much more than others derivatives, i.e.,
\[\nabla _{Z} \equiv \frac{\partial }{\partial Z} \gg \frac{\partial }{\partial X}
,\; \frac{\partial }{\partial Y} \]

Then the system of equations (\ref{eq12})-(\ref{eq13}) is simplified and takes the following form:

\begin{equation} \label{eq15} \partial _{T} W_{1} -\nabla _{{ Z}}^{2}
W_{1} +\nabla _{Z} \left(\overline{v_{0}^{z} v_{0}^{x} }\right)=0, \quad  W_{-1}^{x}
=W_{1}  \end{equation}

\begin{equation} \label{eq16} \partial _{T} W_{2} -\nabla _{{ Z}}^{2}
W_{2} +\nabla _{Z} \left(\overline{v_{0}^{z} v_{0}^{y} }\right)=0, \quad W_{-1}^{y} =W_{2}  \end{equation}

\begin{equation}
\label{eq17} \partial _{T} T_{-1} -Pr^{-1} { \Delta T}_{-1} +\nabla _{{
Z}} \left(\overline{{ v}_{0}^{{ z}} { T}_{0} }\right)=0 \end{equation}

Equations (\ref{eq15})-(\ref{eq16}) describe the evolution of large-scale eddy fields $\vec{W}$ . In order to obtain the final closed form of equations (\ref{eq15})-(\ref{eq16}) we have to calculate the Reynolds stresses $\nabla _{k} \left(\overline{v_{0}^{k} v_{0}^{i} }\right)$. This shows that we need to find solutions for the small-scale velocity field  $\vec{v}_{0} $. Appendix II contains a detailed technique to calculate the velocity field in a rotating stratified medium. Further, in Appendix III solutions for small-scale velocity field $\vec{v}_{0} $ are used to find the Reynolds stresses. Then equations (\ref{eq15})-(\ref{eq16}) take a closed form:
\begin{equation}\label{eq18}
(\partial _T  - \nabla _Z^2 )\widetilde W_1  = \frac{{f_0^2 }}{2}D_2 \nabla _Z \left[ {\frac{{1 + \widetilde W_2^2  - Ra}}{{(1 + \widetilde W_2^2 )((1 + \widetilde W_2^2 )^2  + 2(D_2^2  - Ra)(1 - \widetilde W_2^2 ) + (D_2^2  - Ra)^2 )}}} \right] \end{equation}
\begin{equation}\label{eq19}
(\partial _T  - \nabla _Z^2 )\widetilde W_2  = -\frac{{f_0^2 }}{2}D_1 \nabla _Z \left[ {\frac{{1 + \widetilde W_1^2  - Ra}}{{(1 + \widetilde W_1^2 )((1 + \widetilde W_1^2 )^2  + 2(D_1^2  - Ra)(1 - \widetilde W_1^2 ) + (D_1^2  - Ra)^2 )}}} \right] \end{equation}

To simplify  the equations, we use here the designations: $\widetilde{W}_{1} =1-W_{1} $ , $\widetilde{W}_{2} =1-W_{2} $. Thus, in this section we obtain the closed equations (\ref{eq18})-(\ref{eq19}), which will be called the equations of nonlinear vortex dynamo in obliquely rotating stratified fluids with small scale non-helical force. If the rotation effect disappears ($\Omega =0$ ), then the usual diffusion dissipation of large-scale fields occurs. In the limit of a homogeneous fluid the equation (\ref{eq18}), (\ref{eq19}) coincide with the results of \cite{21s}. We consider first the stability of small perturbations of fields (linear theory) and then examine the question of the possible existence of stationary structures.

\section{Large-scale instability}

Equations (\ref{eq18})-(\ref{eq19}) describe the nonlinear dynamics of large scale disturbances of the vortex field $\vec{W}=\left(W_{1} ,W_{2} \right)$. Therefore it is interesting to clarify the question of the stability of small perturbations of the field  $\vec{W}$. Then for small values $\left(W_{1} ,W_{2} \right)$  the equations (\ref{eq18})-(\ref{eq19}) are linearized and can be reduced to the following system of linear equations:

\begin{equation} \label{eq20}
 \left\{\begin{array}{c} {\partial _{T} W_{1}
-\nabla _{Z}^{2} W_{1} -\alpha _{2} \nabla _{Z} W_{2} =0} \\ {\partial _{T} W_{2}
-\nabla _{Z}^{2} W_{2} +\alpha _{1} \nabla _{Z} W_{1} =0} \end{array}\right.
\end{equation}
where we have introduced the following designations  for the coefficients:

 \[\alpha _1  = f_0^2 D_1 \left[ {\frac{{(D_1^2  - Ra - 2)(2 - Ra) + Ra(4 + (D_1^2  - Ra)^2 )}}{{(4 + (D_1^2  - Ra)^2 )^2 }}} \right] \]
\begin{equation} \label{eq21}\end{equation}
\[\alpha _2  = f_0^2 D_2 \left[ {\frac{{(D_2^2  - Ra - 2)(2 - Ra) + Ra(4 + (D_2^2  - Ra)^2 )}}{{(4 + (D_2^2  - Ra)^2 )^2 }}} \right] \]

\begin{figure}
  \centering
  \includegraphics[height=7 cm]{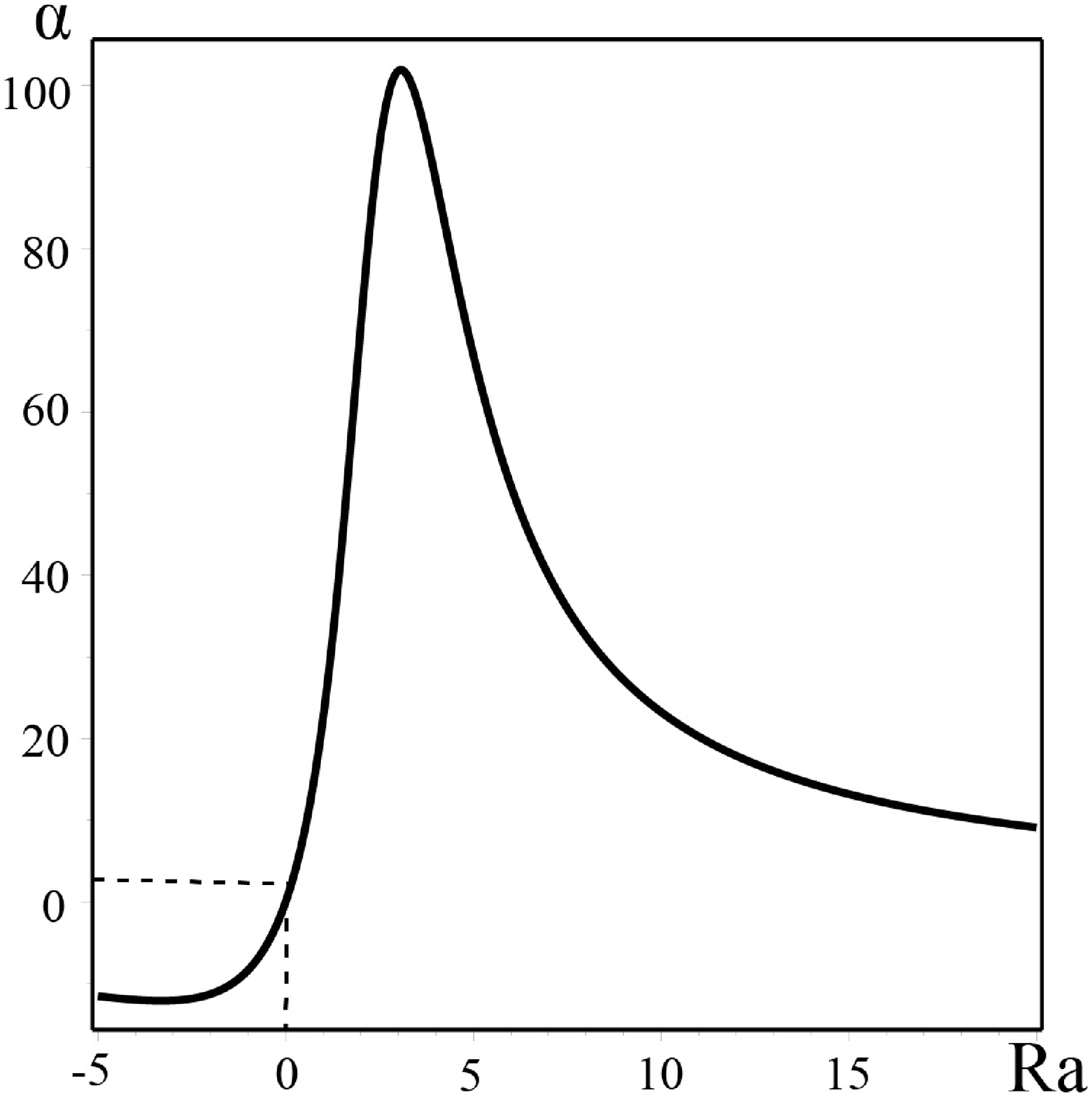}
  \includegraphics[height=7 cm]{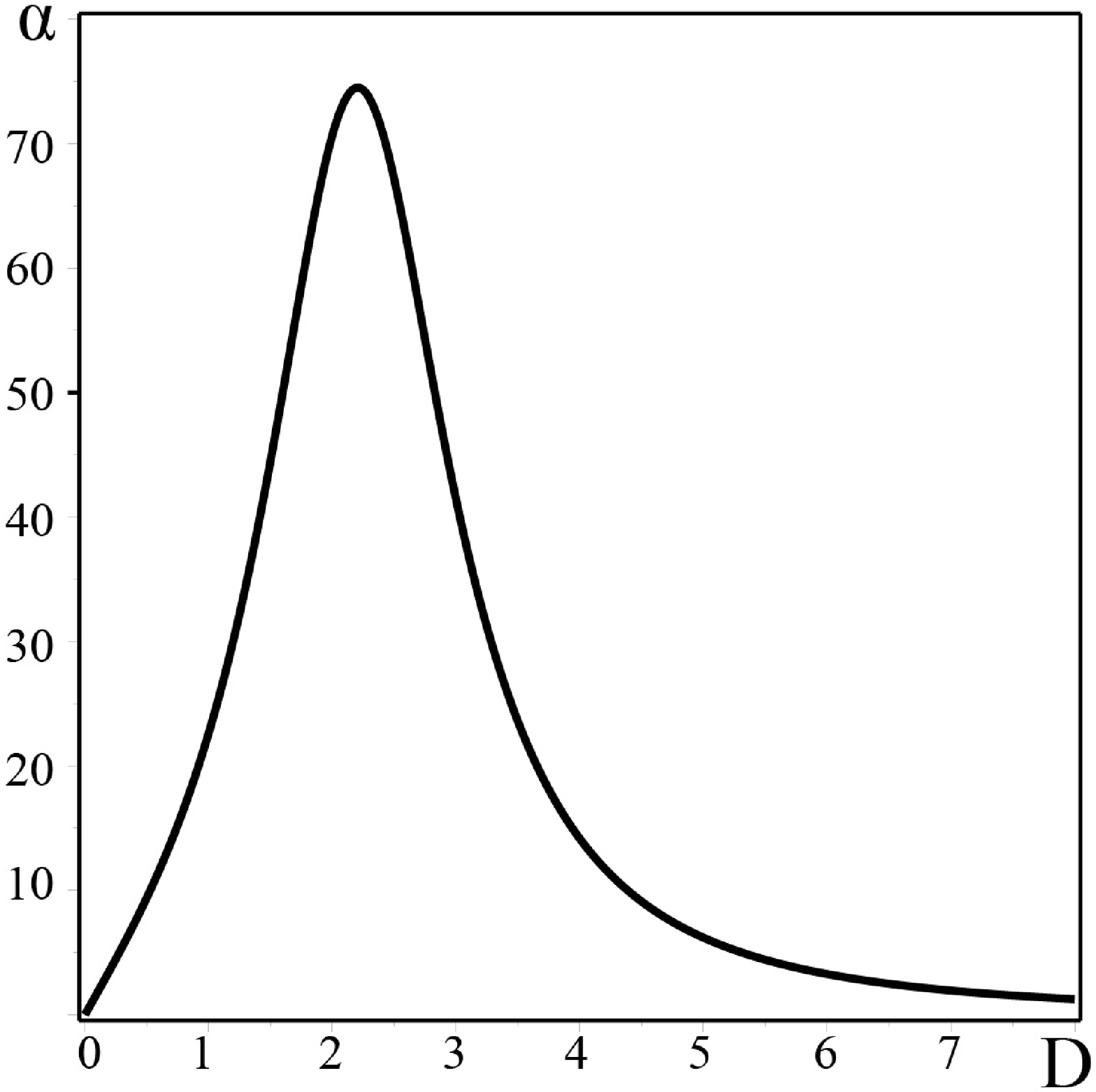}\\
  \caption{On the left the plot of $\alpha $ -- effect of parameter stratification of the medium $Ra$ (Rayleigh number), and on the right the plot of the $\alpha $ -- effect of the parameter of rotation  $D$.} \label{fg3}
\end{figure}

It is clear that equations (\ref{eq20}) are similar to the equations for the vortex dynamo \cite{9s}-\cite{15s}. To study the large-scale instability described by the system of equations (\ref{eq20}), we choose perturbations in the form of plane waves with wave vector $\vec{K}{\rm \parallel }OZ$, i.e.

\begin{equation} \label{eq22}
W_{1,2}  = A_{W_{1,2}}\exp ( -i\omega t + iKZ)
\end{equation}
Substituting (\ref{eq22}) into the system of equations (\ref{eq20}) we get the dispersion equation:
\begin{equation} \label{eq23}
 \left(-i\omega +K^{2} \right)^{2} -\alpha _{1} \alpha _{2} K^{2} =0
\end{equation}
From equation (\ref{eq23}) we find the increment of instability:
\begin{equation} \label{eq24}
 \Gamma =Im \omega  =\pm \sqrt{ \alpha_{1} \alpha_{2} } K-K^{2}
\end{equation}

Solutions (\ref{eq24}) show the existence of instabilities for large-scale vortical perturbations when $\alpha _{1} \alpha _{2} >0$. If $\alpha _{1} \alpha _{2} <0$, damped oscillations arise with frequency $\omega _{0} =\sqrt{\alpha _{1} \alpha _{2} } K$ instead of the instabilities. The coefficients  $\alpha _{1} $,  $\alpha _{2} $  give a positive feedback loop between the components of the velocity. It should be noted that in the linear theory, the coefficients $\alpha _{1} $, $\alpha _{2} $  do not depend on the amplitudes of the fields and depend only on the rotation parameters  $D_{1,2}$, Rayleigh number  $Ra$  and amplitude of the external force $f_{0} $ . Let us analyze the dependence of these coefficients on the dimensionless parameters assuming, for simplicity, that the dimensionless amplitude of the external force $f_{0} $ is equal $f_{0} =10$.  In the coefficients $\alpha _{1} $,  $\alpha _{2} $ , instead of the Cartesian projections $D_{1} $ and $D_{2} $ it is convenient to use their projections in  the spherical coordinate system  $(D,\phi ,\theta )$ (see Fig. \ref{fg2}). The coordinate surface $D=\textrm{const}$ is a sphere, $\theta $ is latitude $\theta \in [0, \pi ]$ , $\phi $ is longitude  $\phi \in [0, 2\pi ]$.

\begin{figure}
  \centering
  \includegraphics[height=7 cm]{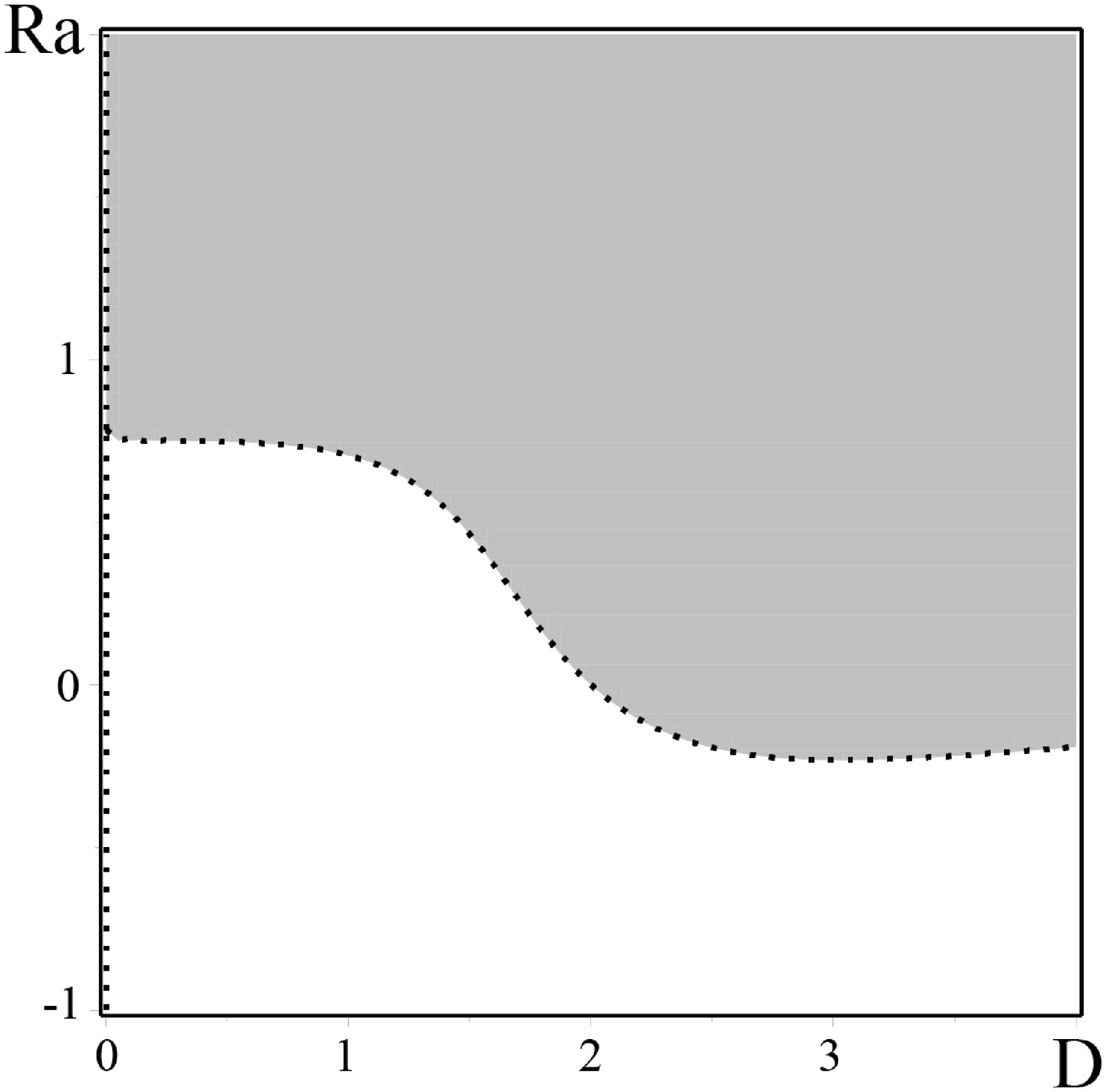}
  \includegraphics[height=7 cm]{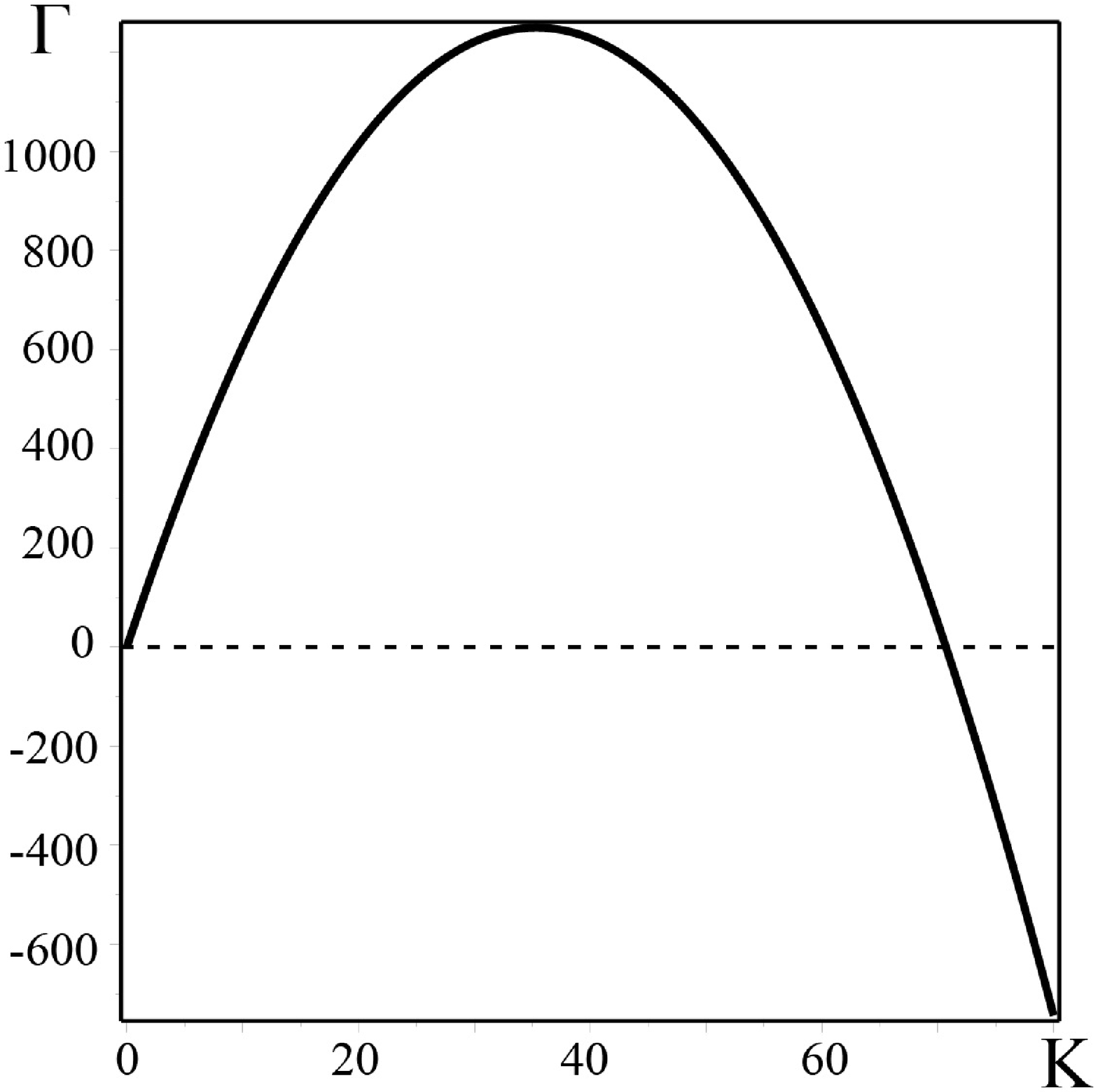}\\
  \caption {On the left the plot for $\alpha $ in the plane $(D,Ra)$, where the gray color shows the region corresponding to positive values $\alpha $ (unstable solutions), and the white negative values $\alpha $. On the right is the plot of the dependence of the instability increment on the wave numbers $K$ for the parameters $D=2$, $Ra=2$.} \label{fg4}
\end{figure}

We analyze the dependence of the amplification coefficients $\alpha _{1} $, $\alpha _{2} $ on the effects of rotation and stratification, assuming for simplicity that $D_{1} =D_{2} $, which corresponds to a fixed value of longitude $\phi =\pi /4+\pi n$, where $n=0,1,2...k$, $k$ is integer. In this case, the amplification coefficients of the vortex perturbations are, respectively, equal to:

\[\alpha =\alpha _{1} =\alpha _{2}  = f_0^2 \sqrt 2 D\sin \theta \left[ {\frac{{4(D^2 \sin ^2 \theta  - 2Ra - 4)(2 - Ra) + 2Ra(16 + (D^2 \sin ^2 \theta  - 2Ra)^2 )}}{{(16 + (D^2 \sin ^2 \theta  - 2Ra)^2 )^2 }}} \right] \]

We can see from this equation that at the poles  $(\theta =0,\; \theta =\pi )$  the generation of vortex perturbations is not effective because $\alpha \to 0$. The dependence of $\alpha $  coefficient on the stratification parameter of fluid (Rayleigh number $Ra$) at a fixed latitude  $\theta=\pi/2$ and the number $D=2$ is presented  in the left part of Fig. \ref{fg3}. Also it shows the case of a homogeneous medium $Ra=0$, where the generation of large-scale vortex perturbations is caused by the external non-helical small-scale force and the Coriolis force \cite{21s}. Fig. \ref{fg3} shows that the presence of temperature stratification ( $Ra\ne 0$) can engender a significant increase the coefficient $\alpha $ . Consequently we have faster generation of large-scale vortex perturbations than in homogeneous medium. This effect m anifestes especially with  numbers $Ra\to 2$. Further, with the increasing of Rayleigh numbers, the value of coefficient $\alpha $ decreases. It is also interesting to find out the impact  of rotation effect  on the amplification coefficients $\alpha $ . For these purposes, we take the value of the Rayleigh number  $Ra=2$  at $\theta =\pi /2$. For this case the functional dependency $\alpha (D)$  is shown in the right part of Fig. \ref{fg3}. This shows that for some parameter of $D$ the coefficent $\alpha $ reaches its maximum value $\alpha _{\max } $. Then with the increasing of $D$ the coefficient $\alpha $ tends gradually to zero  i.e. the suppression of $\alpha $ -- effect occurs. A similar phenomenon was described in \cite{19s}, \cite{20s}. The left part of Fig. \ref{fg4} shows the plot of the joint effect of rotation and stratification in the plane $(D,Ra)$ . Here the instability area is highlighted in gray. The maximum increment of instability $\Gamma _{max} =\frac{\alpha _{1} \alpha _{2} }{4} $  is reached on the wave number $K_{max} =\frac{\sqrt{\alpha _{1} \alpha _{2} } }{2} $ . The plot of the function $\Gamma $ from the wave number $K$ (see right part of Fig. \ref{fg4})  has a standard form of $\alpha $ -- effect. Thus, the development of large-scale instability in rotating stratified atmosphere generates the large-scale helical circularly polarized vortices of Beltramy type.

\begin{figure}
  \centering
  \includegraphics[width=6 cm]{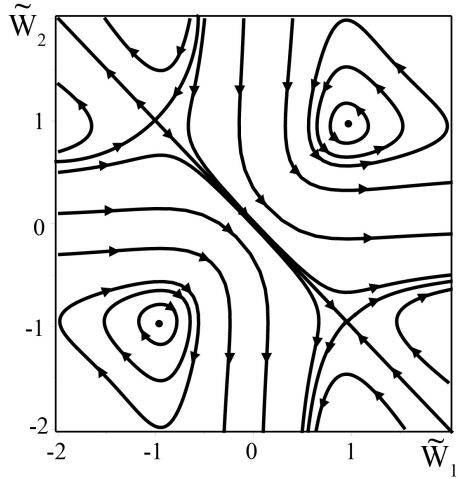}\\
  \caption {The phase plane of  the dynamical system equations  (\ref{eq25})-(\ref{eq26}) with $C_{1} =-1$ and  $C_{2} =1$. One can see the presence of closed trajectories around the elliptic points and  separatrices which connect the hyperbolic points.}\label{fg5}
\end{figure}

\section{Stationary nonlinear vortex structures}

It is obvious that with the increase of amplitude, nonlinear terms decrease and the instability becomes saturated. As a  result the nonlinear vortex structures appear. In order to find these structures let us examine the stationary case of equations  (\ref{eq18})-(\ref{eq19}) and intergrate once with respect to $Z$. For the sake of simplicity we assume that $D_1=D_2$ and $\theta=\pi/2$. Consequently we get a system of nonlinear equations of the following form:

\begin{figure}
  \centering
    \includegraphics[width=6 cm]{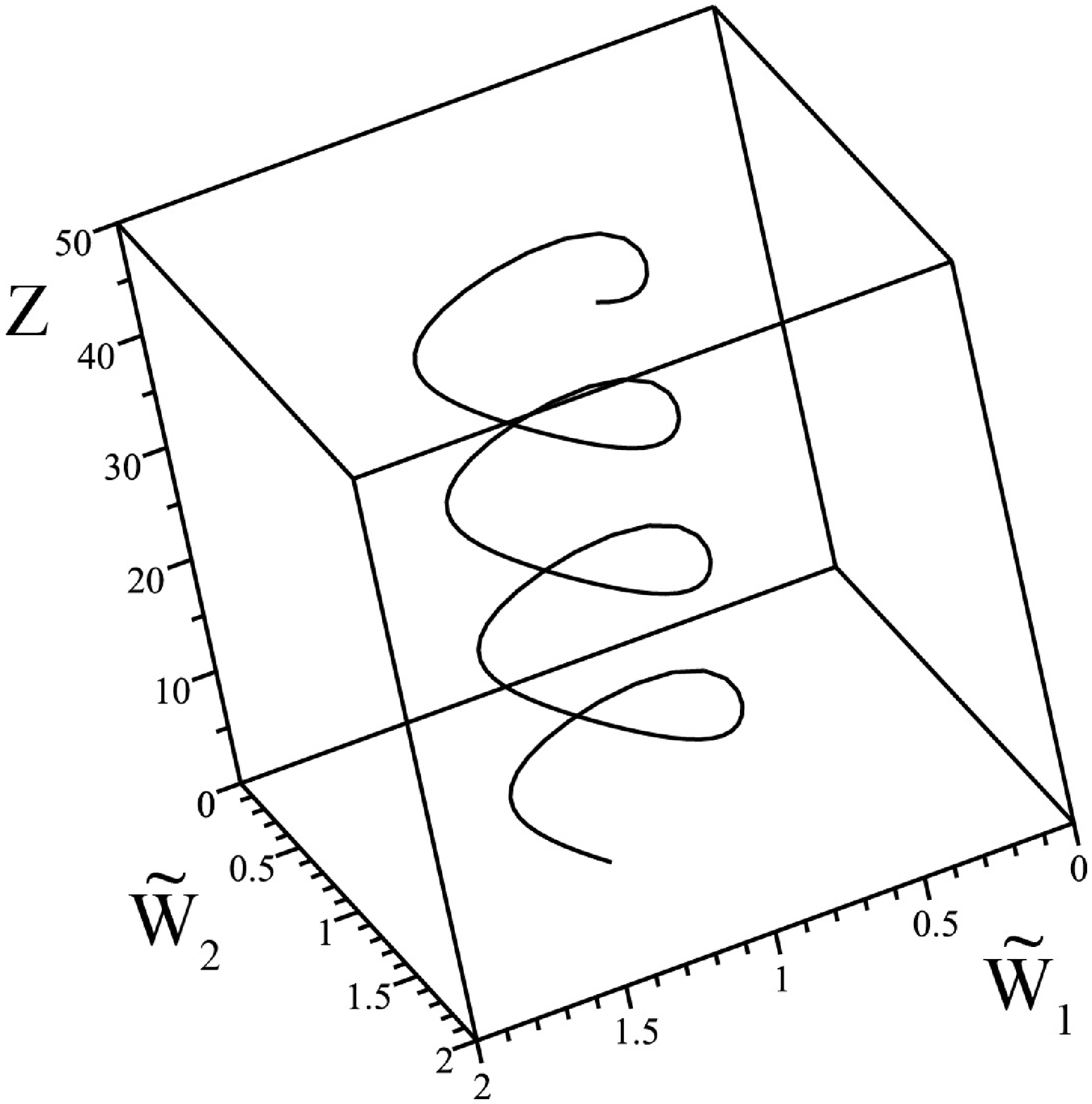}
		\includegraphics[width=6 cm]{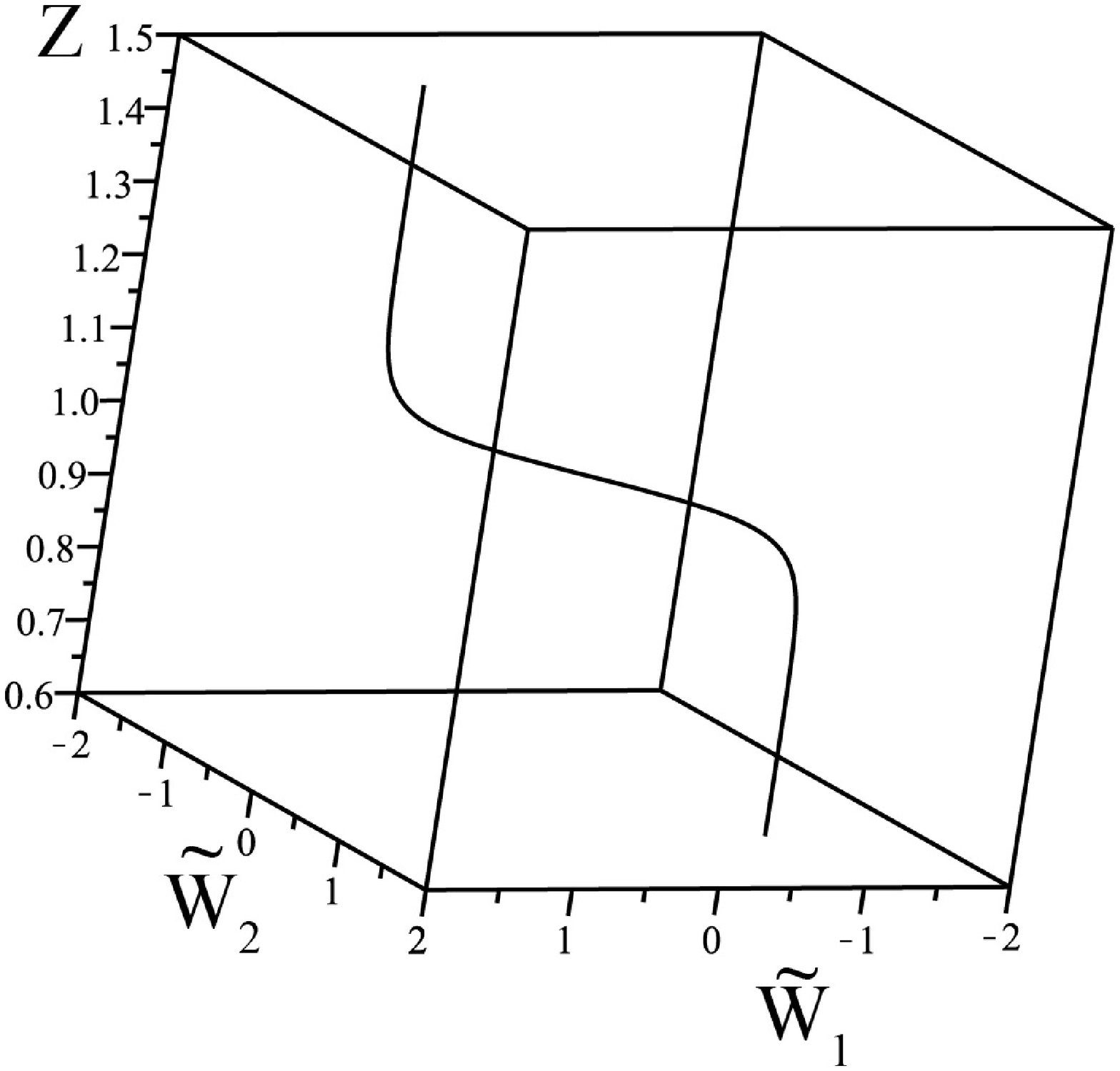}\\
  \caption{On the left a nonlinear helical wave, which corresponds to a closed trajectory on the phase plane; on the right a localized nonlinear vortex structure (kink), which corresponds to the separatrix on the phase plane ($C_{1} =-1$, $C_{2} =1$, $D=Ra=2$)}\label{fg6}
\end{figure}

\begin{equation}\label{eq25}
\frac{{d\widetilde W_1 }}{{dZ}} =  - f_0^2 D\sqrt 2 \frac{{1 + \widetilde W_2^2  - Ra}}{{(1 + \widetilde W_2^2 )(4(1 + \widetilde W_2^2 )^2  + 4(D^2  - 2Ra)(1 - \widetilde W_2^2 ) + (D^2  - 2Ra)^2 )}} + C_1
\end{equation}

\begin{equation}\label{eq26}
\frac{{d\widetilde W_2 }}{{dZ}} =   f_0^2 D\sqrt 2 \frac{{1 + \widetilde W_1^2  - Ra}}{{(1 + \widetilde W_1^2 )(4(1 + \widetilde W_1^2 )^2  + 4(D^2  - 2Ra)(1 - \widetilde W_1^2 ) + (D^2  - 2Ra)^2 )}} + C_2
\end{equation}

Here,  $C_{1},\; C_{2} $ are arbitrary constants of integration. It should be noted that the dynamic system of equations (\ref{eq25})-(\ref{eq26}) is conservative, and hence is Hamiltonian. It's easy to find it we write down the equations (\ref{eq25})-(\ref{eq26}) in the Hamiltonian form:
\[\frac{{d\widetilde W_1 }}{{dZ}} =  - \frac{{\partial H}}{{\partial \widetilde W_2 }}, \quad \frac{{d\widetilde W_2 }}{{dZ}} =   \frac{{\partial H}}{{\partial \widetilde W_1 }} \]
where the Hamiltonian has the form:
\begin{equation}\label{eq27}
H = H_1 (\widetilde W_1 ) + H_2 (\widetilde W_2 ) + C_2 \widetilde W_1  - C_1 \widetilde W_2
\end{equation}

The functions  $H_{1,2} $   are respectively equal to:

\[H_{1,2}  = f_0^2 D\sqrt 2 \int {\frac{{(1 + \widetilde W_{1,2}^2  - Ra)d\widetilde W_{1,2} }}{{(1 + \widetilde W_{1,2}^2 )(4(1 + \widetilde W_{1,2}^2 )^2  + 4(D^2  - 2Ra)(1 - \widetilde W_{1,2}^2 ) + (D^2  - 2Ra)^2 )}}}\]

Let us put  $D=Ra=2$ and $f_{0} =10$. Then we  can calculate the Hamiltonian (\ref{eq27}):

\[H =  - \frac{{25}}{2}\sqrt 2 \left( {\frac{{\widetilde W_1 (\widetilde W_1^2  + 3)}}{{(\widetilde W_1^2  + 1)^2 }} + \frac{{\widetilde W_2 (\widetilde W_2^2  + 3)}}{{(\widetilde W_2^2  + 1)^2 }} + arctg\widetilde W_1  + arctg\widetilde W_2 } \right) + C_2 \widetilde W_1  - C_1 \widetilde W_2 \]

Since the equations (\ref{eq25})-(\ref{eq26})  are Hamiltonian, only fixed points of  two types: elliptic and hyperbolic can be observed in a phase space. This can be checked if we carry out a qualitative analysis of the system of equations (\ref{eq25})-(\ref{eq26}). Linearizing the right-hand sides of equations (\ref{eq25})-(\ref{eq26}) in the neighborhood of fixed points, we establish their type and construct a phase portrait. As a result of the analysis, we find the appearance of four fixed points, two of hyperbolic and two of elliptic type. Phase portrait of a dynamical system of equations (\ref{eq25})-(\ref{eq26}) for the parameters  $C_{1} =-1$, $C_{2} =1$,  $D=Ra=2$  and  $f_{0} =10$ is shown in Fig. \ref{fg5}. The phase portrait allows us to describe qualitatively the possible stationary solutions. The most interesting localized solutions correspond to the phase portrait trajectories, which connect the stationary (singular) points on the phase plane. Fig. \ref{fg5} presents closed trajectories on the phase plane around the elliptic points and separatrices which connect the hyperbolic points. Closed trajectories correspond to nonlinear periodic solutions or nonlinear waves.  The separatrices correspond  to a localized vortex structures of the kinks type (see Fig.\ref{fg6} ).

\section{Conclusion}

In this paper we obtain a new type of large-scale instability generated by vertical temperature gradient and small-scale force with zero helicity $\vec{F}_{0} rot\vec{F}_{0} =0$ in an inclined rotating fluid. This force supports small-scale fluctuations in the fluid and models the effect of small-scale turbulence with the Reynolds number $R \ll1$. It is assumed that the external force is in the plane $(X,Y)$. The force of gravity is directed vertically downwards along the axis  $OZ$. Using the method of multiscale asymptotic expansions we get in our work the closed system of equations for large scale perturbations of velocity. For small amplitudes, this system of equations describes the instability, which is called the hydrodynamic $\alpha $ -- effect, since there is a positive feedback between the components of the velocity. It is shown that the instability arises only in the case, when the angular velocity vector of rotation deviates from the axis $OZ$. Joint effect of rotation and stratification of the medium (temperature heated from below) leads to a substantial enhancement of large-scale vortex perturbations, unlike the case of a homogeneous medium \cite{21s}. This phenomenon appears especially when the parameters of medium $D \to 2$,  $Ra\to 2$  (see Fig. \ref{fg3} ). In this case there is a maximum generation of small-scale helical motions due to the action of the Coriolis force and the inhomogeneity of the medium temperature. The rapid growth of vortex perturbations contributes to the increasing role of nonlinear effects and the consequent saturation of the instability. The study of stationary state by numerical method with parameters $D=Ra=2$ shows the existence of two types of solutions: nonlinear waves and kinks. These solutions are similar to those found in homogeneous obliquely rotating fluid \cite{21s}.

\section*{Appendix I. Multiscale asymptotic developments}

Let us find the algebraic structure of the asymptotic development in various orders of $R$ , starting from the lowest one. In order of $R^{-3} $ there is only one equation:
\begin{equation}\label{eqI1} \partial _{i} P_{-3} =0\Rightarrow \; P_{-3} =P_{-3} \left(X\right) \end{equation}
In order $R^{-2}$ appears the equations:
\begin{equation}\label{eqI2}\partial _{i} P_{-2} =0\; \; \; \; \Rightarrow \; \; P_{-2} =P_{-2} \left(X\right) \end{equation}
In order  $R^{-1}$, we obtain more complicated system of equations:
\begin{equation}\label{eqI3} \partial _{t} W_{-1}^{i} +W_{-1}^{k} \partial _{k} W_{-1}^{i} =-\partial _{i} P_{-1}
-\nabla _{i} P_{-3} +\partial _{k}^{2} W_{-1}^{i} +\varepsilon _{ijk} W_{j} D_{k}
e_{k} +\widetilde{Ra}e_{i} T_{-1}  \end{equation}

  \begin{equation}\label{eqI4} \partial
_{t} T_{-1} -Pr^{-1} \partial _{k}^{2} T_{-1} =-W_{-1}^{k} \partial _{k} T_{-1} + W_{-1}^{z} \end{equation}

\begin{equation}\label{eqI5} \partial _{i} W_{-1}^{i} =0 \end{equation}
The averaging of equations (\ref{eqI3})-(\ref{eqI5}) over the fast variables give the following secular equations:
\begin{equation}\label{eqI6}-\nabla _{i} P_{-3} +\widetilde{Ra}e_{i} T_{-1} +\varepsilon _{ijk} W_{j} D_{k} =0 \end{equation}

\begin{equation}\label{eqI7} W_{-1}^{z} =0 \end{equation}

In zero order in $R^{0} $ we have the equations:
\begin{equation}\label{eqI8} \partial _{t} v_{0}^{i} +W_{-1}^{k} \partial _{k} v_{0}^{i} +v_{0}^{k} \partial
_{k} W_{-1}^{i} =-\partial _{i} P_{0} -\nabla _{i} P_{-2} +\partial _{k}^{2} v_{0}^{i}
+ \varepsilon _{ijk} v_{0}^{j} D_{k} +\widetilde{Ra}e_{i} T_{0} +F_{0}^{i} \end{equation}

\begin{equation}\label{eqI9}\partial _{t} T_{0} -Pr^{-1} \partial _{k}^{2} T_{0} =-W_{-1}^{k} \partial _{k}
T_{0}-\partial _{k}(v_{0}^{k} T_{-1}) +v_{0}^{z} \end{equation}

\begin{equation}\label{eqI10}\partial
_{i} v_{0}^{i} =0 \end{equation}

These equations give one secular equation:
\begin{equation}\label{eqI11}
  \nabla P_{-2} =0\; \; \; \; \; \; \Rightarrow \; \; \; P_{-2} =const
\end{equation}
Let us consider the equations of the first approximation $R^{1} $:

\[ \partial _{t} v_{1}^{i} +W_{-1}^{k} \partial _{k} v_{1}^{i}
+v_{0}^{k} \partial _{k} v_{0}^{i} +v_{1}^{k} \partial _{k} W_{-1}^{i} +W_{-1}^{k}
\nabla _{k} W_{-1}^{i} =-\nabla _{i} P_{-1} -\partial _{i} \left(P_{1} +\overline{P}_{1}
\right)+\partial _{k}^{2} v_{1}^{i} +\]
\begin{equation}\label{eqI12}
{+2\partial _{k} \nabla _{k} W_{-1}^{i}
+Rae_{i}T_{1}+\varepsilon _{ijk} v_{1}^{j} D_{k} }  \end{equation}

\begin{equation}\label{eqI13} \partial
_{t} T_{1} -Pr^{-1} \partial _{k}^{2} T_{1} -Pr^{-1}2\partial _{k} \nabla _{k} T_{-1}
=-W_{-1}^{k} \partial _{k} T_{1} -W_{-1}^{k} \nabla _{k} T_{-1}
-v_{0}^{k} \partial _{k} T_{0} -v_{1}^{k} \partial _{k} T_{-1} +v_{1}^{z} \end{equation}

 \begin{equation}\label{eqI14} \partial
_{i} v_{1}^{i} +\nabla _{i} W_{-1}^{i} =0 \end{equation}

The secular equations follow from this system of equations:

 \begin{equation}\label{eqI15} W_{-1}^{k} \nabla _{k} W_{-1}^{i} =-\nabla _{i} P_{-1} \end{equation}

\begin{equation}\label{eqI16} W_{-1}^{k}
\nabla _{k} T_{-1} =0 \end{equation}                                                                                                                      \begin{equation}\label{eqI17} \nabla
_{i} W_{-1}^{i} =0 \end{equation}                                                                                                                      The secular equation (\ref{eqI15})-(\ref{eqI17}) are satisfied by choosing the following geometry:
\begin{equation}\label{eqI18}\vec{W}_{-1} =\left(W_{-1}^{x} \left(Z\right),W_{-1}^{y} \left(Z\right),0\right),T_{-1} =T_{-1} \left(Z\right),P_{-1} =\textrm{const} \end{equation}

In the second order   $R^{2} $ , we obtain the equations:

 \[\partial _{t} v_{2}^{i} +W_{-1}^{k} \partial _{k} v_{2}^{i} +v_{0}^{k}
\partial _{k} v_{1}^{i} +W_{-1}^{k} \nabla _{k} v_{0}^{i} +v_{0}^{k} \nabla _{k}
W_{-1}^{i} +v_{1}^{k} \partial _{k} v_{0}^{i} +v_{2}^{k} \partial _{k} W_{-1}^{i}
=-\nabla _{i} P_{2} -\nabla _{i} P_{0} +\]
\begin{equation}\label{eqI19}
{+\partial _{k}^{2} v_{2}^{i} +2\partial
_{k} \nabla _{k} v_{0}^{i} +\widetilde{Ra}e_{i} T_{2} +{\varepsilon }_{ijk} {v}_{2}^{j}
{D}_{k} }  \end{equation}

\[\partial _{t} T_{2} -Pr^{-1} \partial _{k}^{2} T_{2} -Pr^{-1} 2\partial _{k} \nabla
_{k} T_{0} =-W_{-1}^{k} \partial _{k} T_{2} -W_{-1}^{k} \nabla _{k} T_{0} -v_{0}^{k}
\partial _{k} T_{1} -v_{0}^{k} \nabla _{k} T_{-1}-\]
\begin{equation}\label{eqI20} {-v_{1}^{k} \partial _{k}
T_{0} -v_{2}^{k} \partial _{k} T_{-1} +v_{2}^{z} }  \end{equation}                                                                                                                                                                                                                                       \begin{equation}\label{eqI21} \partial_{i} v_{2}^{i} +\nabla _{i} v_{0}^{i} =0 \end{equation}

It is easy to see that there are no secular terms in this order. Let us consider now the most important order $R^{3} $. In this order we obtain the equations:
 \[\partial _{t} v_{3}^{i} +\partial _{T} W_{-1}^{i} +W_{-1}^{k}
\partial _{k} v_{3}^{i} +v_{0}^{k} \partial _{k} v_{2}^{i} +W_{-1}^{k} \nabla _{k}
v_{1}^{i} +v_{0}^{k} \nabla _{k} v_{0}^{i} +v_{1}^{k} \partial _{k} v_{1}^{i} +v_{1}^{k}
\nabla _{k} W_{-1}^{i} +\]
 \[+v_{2}^{k} \partial _{k} v_{0}^{i}+v_{3}^{k} \partial _{k} W_{-1}^{i}=-\partial _{i}
P_{3} -\nabla _{i} \left(P_{1} +\overline{P}_{1} \right)+\partial _{k}^{2} v_{3}^{i}
+2\partial _{k} \nabla _{k} v_{1}^{i} +\Delta W_{-1}^{i} +\widetilde{Ra}e_{i} T_{3}
 +\]
\begin{equation}\label{eqI22} {+{\varepsilon }_{ijk} {v}_{3}^{j}
{D}_{k} }  \end{equation}

\[\partial _{t} T_{3} +\partial _{T} T_{-1} -Pr^{-1} \partial _{k}^{2}
T_{3} -Pr^{-1} 2\partial _{k} \nabla _{k} T_{1} -Pr^{-1} \Delta T_{-1} =-W_{-1}^{k}
\partial _{k} T_{3} -W_{-1}^{k} \nabla _{k} T_{1} -\]
\begin{equation}\label{eqI23} {-v_{0}^{k} \partial _{k} T_{2} -v_{0}^{k} \nabla _{k} T_{0} -v_{1}^{k} \nabla _{k} T_{1} -v_{1}^{k} \nabla_{k} T_{-1} -v_{2}^{k} \partial _{k} T_{0} -v_{3}^{k} \partial _{k} T_{-1} +v_{3}^{z}
} \end{equation}

\begin{equation}\label{eqI24}\partial _{i} v_{3}^{i} +\nabla _{i} v_{1}^{i} =0 \end{equation}

After averaging this system of equations over the fast variables, we obtain the main system of secular equations to describe the evolution of large-scale perturbations:
 \begin{equation}\label{eqI25} \partial _{T} W_{-1}^{i} -\Delta W_{-1}^{i} +\nabla _{k} \left(\overline{v_{0}^{k}
v_{0}^{i} }\right)=-\nabla _{i} \overline{P}_{1} \end{equation}

 \begin{equation}\label{eqI26}\partial
_{T} T_{-1} -Pr^{-1} \Delta T_{-1} =-\nabla _{k} \left(\overline{v_{0}^{k} T_{0}
}\right) \end{equation}

\section*{Appendix II. Small-scale fields in the zero order in $R$}

In Appendix I, we obtain the equations in the zero order in $R$, which can be written in the following form:

\begin{equation} \label{eqII1}
 \widehat{D}_{W} v_{0}^{i} =-\partial _{i} P_{0}+\widetilde{Ra}e_{i} T_{0}
+\varepsilon _{ijk} v_{0}^{j} D_{k} +F_{0}^{i}
\end{equation}

\begin{equation}
\label{eqII2} \widehat{D}_{\theta } T_{0} =e_{k} v_{0}^{k}  \end{equation}

\begin{equation} \label{eqII3}
 \partial _{i} v_{0}^{i} =0
\end{equation}

where we introduce the designations for operators:

\[\widehat{D}_{W} =\partial _{t} -\partial _{k}^{2} +W_{-1}^{k} \partial _{k} ,\;\widehat{D}_{\theta } =\partial _{t} -Pr^{-1}
\partial ^{2} +W_{-1}^{k} \partial _{k}\]

Small-scale oscillations of temperature are easily found from the equation (\ref{eqII2})

\begin{equation} \label{eqII4} T_{0} =\frac{v_{0}^{z} }{\widehat{D}_{\theta} }  \end{equation}

Let us substitute (\ref{eqII4}) into (\ref{eqII1}) and using the condition of solenoidality of fields $\vec{v}_{0} $ ,$\vec{F}_{0} $ , we can find the pressure $P_{0} $:

\begin{equation} \label{eqII5}
 P_{0} =\widehat{P}_{1} u_{0} +\widehat{P}_{2}v_{0} +\widehat{P}_{3} w_{0}
\end{equation}

Here we introduced the designation for operators

\[\widehat{P}_{1} =\frac{D_{2} \partial _{z}
-D_{3} \partial _{y} }{\partial ^{2} },\quad \widehat{P}_{2} =\frac{D_{3} \partial _{x} -D_{1}
\partial _{z} }{\partial ^{2}}  ,\quad \widehat{P}_{3} =\frac{D_{1} \partial _{y} -D_{2} \partial
_{x} }{\partial ^{2}}+\widetilde{Ra} \frac{{\partial _z }}{{\hat D_\theta \partial^2 }} \]

and velocities: $v_{0}^{x} =u_{0} $, $v_{0}^{y} =v_{0} $, $v_{0}^{z} =w_{0} $.
Using the representation (\ref{eqII5}), we can eliminate pressure from the equations  (\ref{eqII1}) and obtain the system of equations for  velocity fields of the zero approximation:

\[\left(\widehat{D}_{W} +\widehat{p}_{1x} \right)u_{0} +\left(\widehat{p}_{2x} -D_{3}
\right)v_{0} +\left(\widehat{p}_{3x} +D_{2} \right)w_{0} =F_{0}^{x} \]

\begin{equation}\label{eqII6}
\left(D_{3}
+\widehat{p}_{1y} \right)u_{0} +\left(\widehat{D}_{W} +\widehat{p}_{2y} \right)v_{0} +\left(
\widehat{p}_{3y} -D_{1} \right)w_{0} =F_{0}^{y}
\end{equation}

  \[\left(\widehat{p}_{1z} -D_{2} \right)u_{0} +\left(\widehat{p}_{2z} +D_{1} \right)v_{0}
+\left(\widehat{D}_{W} -\frac{\widetilde{Ra}}{\widehat{D}_{\theta}} +\widehat{p}_{3z} \right)w_{0} =0\]
The components of the tensor have the following form:

\[\widehat{p}_{1x} =\frac{D_{2} \partial _{x}
\partial _{z} -D_{3} \partial _{x} \partial _{y} }{\partial ^{2} } ,\widehat{p}_{2x}
=\frac{D_{3} \partial ^{2} _{x} -D_{1} \partial _{x} \partial _{z} }{\partial ^{2}
} ,\widehat{p}_{3x} =\frac{D_{1} \partial _{x} \partial _{y} -D_{2} \partial ^{2}
_{x} }{\partial ^{2} }+\widetilde{Ra} \frac{{\partial _x}{\partial _z }}{{\hat D_\theta \partial^2 }} ,\]
\begin{equation} \label{eqII7}
\widehat{p}_{1y} =\frac{D_{2} \partial _{y} \partial _{z}
-D_{3} \partial ^{2} _{y} }{\partial ^{2} }  , \widehat{p}_{2y} =\frac{D_{3} \partial
_{y} \partial _{x} -D_{1} \partial _{y} \partial _{z} }{\partial ^{2} } ,\widehat{p}_{3y}
=\frac{D_{1} \partial ^{2} _{y} -D_{2} \partial _{y} \partial _{x} }{\partial ^{2}
}+\widetilde{Ra} \frac{{\partial _y}{\partial _z }}{{\hat D_\theta \partial^2 }} ,
\end{equation}
\[ \widehat{p}_{1z} =\frac{D_{2} \partial ^{2} _{z} -D_{3} \partial _{z} \partial
_{y} }{\partial ^{2} } ,\; \widehat{p}_{2z} =\frac{D_{3} \partial _{z} \partial _{x}
-D_{1} \partial ^{2} _{z} }{\partial ^{2} } , \widehat{p}_{3z} =\frac{D_{1} \partial
_{z} \partial _{y} -D_{2} \partial _{z} \partial _{x} }{\partial ^{2}}+\widetilde{Ra} \frac{\partial^{2}_{z}}{{\hat D_\theta \partial^2 }}\]

The solution for equations system (\ref{eqII6}) can be found in accordance with Cramer's rule:

\[u_{0} =\frac{1}{\Delta } \left\{ \left[\left(\widehat{D}_{W} +\widehat{p}_{2y} \right)\left(\widehat{D}_{W} -\frac{\widetilde{Ra}
 }{\widehat{D}_{\theta} } +\widehat{p}_{3z} \right)-
\left(\widehat{p}_{2z} +D_{1} \right)\left(\widehat{p}_{3y} -D_{1} \right)\right]F_{0}^{x}
\right. +  \]

\begin{equation} \label{eqII8}
 \left.{+ \left[\left(\widehat{p}_{3x} +D_{2} \right)\left(\widehat{p}_{2z} +D_{1} \right)-
\left(\widehat{p}_{2x} -D_{3} \right)\left(\widehat{D}_{W} -\frac{\widetilde{Ra}}{\widehat{D}_{\theta} } +\widehat{p}_{3z} \right) \right]F_{0}^{y}
} \right\}
\end{equation}

\[ v_{0} =\frac{1}{\Delta } \left\{ \left[\left(\widehat{D}_{W}+\widehat{p}_{1x} \right)\left(\widehat{D}_{W} -\frac{\widetilde{Ra}
 }{\widehat{D}_{\theta} } +\widehat{p}_{3z} \right)-
\left(\widehat{p}_{3x} +D_{2} \right)\left(\widehat{p}_{1z} -D_{2} \right) \right]F_{0}^{y}
+ \right.\]

\begin{equation} \label{eqII9} \left. {+\left[\left(\widehat{p}_{3y} -D_{1} \right) \left(\widehat{p}_{1z} -D_{2} \right)-
\left(D_{3} +\widehat{p}_{1y} \right)\left(\widehat{D}_{W} -\frac{\widetilde{Ra} }{\widehat{D}_{\theta} } +\widehat{p}_{3z} \right) \right]F_{0}^{x}
} \right\} \end{equation}

\[ w_{0} =\frac{1}{\Delta } \left\{ \left[\left(D_{3} +\widehat{p}_{1y} \right)\left(\widehat{p}_{2z} +D_{1} \right)-\left(
\widehat{D}_{W} +\widehat{p}_{2y} \right)\left(\widehat{p}_{1z} -D_{2} \right) \right]F_{0}^{x} + \right.\]

\begin{equation} \label{eqII10}
\left.{+\left[\left(\widehat{p}_{2x} -D_{3} \right)\left(\widehat{p}_{1z} -D_{2} \right)-
\left(\widehat{D}_{W} +\widehat{p}_{1x} \right)\left(\widehat{p}_{2z} +D_{1} \right)
\right]F_{0}^{y} }\right\}
\end{equation}
Here  $\Delta $ is the determinant of the system of equations (\ref{eqII6}):

\[\Delta =\left(\widehat{D}_{W} +\widehat{p}_{1x} \right)\left(\widehat{D}_{W} +\widehat{p}_{2y} \right)\left(\widehat{D}_{W} -\frac{\widetilde{Ra} }{\widehat{D}_{\theta} } +\widehat{p}_{3z} \right)+\]

\[+\left(D_{3} +\widehat{p}_{1y} \right)\left(\widehat{p}_{2z} +D_{1} \right)\left(
\widehat{p}_{3x} +D_{2} \right)+\left(\widehat{p}_{2x} -D_{3} \right)\left(\widehat{p}_{3y}
-D_{1} \right)\left(\widehat{p}_{1z} -D_{2} \right)-\]

\[-\left(\widehat{p}_{3x}
+D_{2} \right)\left(\widehat{D}_{W} +\widehat{p}_{2y} \right)\left(\widehat{p}_{1z} -D_{2}
\right)-\]

\[-\left(\widehat{p}_{2z} +D_{1} \right)\left(\widehat{p}_{3y} -D_{1}
\right)\left(\widehat{D}_{W}  +\widehat{p}_{1x} \right)-\]

\begin{equation}\label{eqII11}
    -\left(D_{3} +\widehat{p}_{1y}
\right)\left(\widehat{p}_{2x} -D_{3} \right)\left(\widehat{D}_{W} -\frac{\widetilde{Ra}
 }{\widehat{D}_{\theta} } +\widehat{p}_{3z} \right)
\end{equation}
In order to calculate the expressions (\ref{eqII8})-(\ref{eqII11}) we present the external force  (\ref{eq6})   in complex form:
\begin{equation} \label{eqII12}
 \vec{F}_{0} = \vec{i} \, \frac{f_{0} }{2} \; e^{i
\phi _{2} } +\vec{j} \, \frac{f_{0} }{2} e^{i\phi _{1} } + k.c.
\end{equation}
Then all operators in formulae (\ref{eqII8})-(\ref{eqII11}) act from the left on their eigen function. In particular:
\[\widehat{D}_{W,H} e^{i\phi _{1} } =e^{i\phi
_{1} } \widehat{D}_{W,\theta} \left(\vec{\kappa }_{1} , -\omega _{0} \right), \quad \widehat{D}_{W,\theta}
e^{i\phi _{2} } =e^{i\phi _{2} } \widehat{D}_{W,\theta} \left(\vec{\kappa }_{2} ,- \omega _{0} \right), \]
\begin{equation} \label{eqII13}
 \Delta e^{i\phi _{1} } =e^{i\phi _{1} } \Delta \left(\vec{\kappa
}_{1} ,\; -\omega _{0} \right),\quad \Delta e^{i\phi _{2} } =e^{i\phi _{2} } \Delta \left(
\vec{\kappa }_{2} ,\; -\omega _{0} \right)
\end{equation}
To simplify the formulae, let us choose    $\kappa _{0} =1$, $\omega _{0} =1$    and introduce new designations:
\begin{equation}\label{eqII14}
\widehat{D}_{W} \left(\vec{\kappa }_{1} ,\; -\omega _{0} \right)=
\widehat{D}_{W_{1} }^{*} =1-i\left(1-W_{1} \right),\quad \widehat{D}_{W} \left(\vec{\kappa
}_{2} ,\; -\omega _{0} \right)=\widehat{D}_{W_{2} }^{*} =1-i\left(1-W_{2} \right) \end{equation}

\[\widehat{D}_{\theta}
\left(\vec{\kappa }_{1} ,\; -\omega _{0} \right)=\widehat{D}_{\theta_{1} }^{*} =Pr^{-1}
-i\left(1-W_{1} \right),\quad \widehat{D}_{\theta} \left(\vec{\kappa }_{2} ,\; -\omega _{0}
\right)=\widehat{D}_{\theta_{2} }^{*} =Pr^{-1} -i\left(1-W_{2} \right)\]

Complex-conjugate quantities  will be denoted with asterisk. When performing further calculations, the part of component tensors   $\widehat{p}_{ij} \left(\vec{\kappa }_{1} \right)$ and  $\widehat{p}_{ij} \left(\vec{\kappa }_{2} \right)$ vanishes. Taking this into account the velocity field of  zero approximation has the following form:
\begin{equation}\label{eqII15}
 u_{0} =\frac{f_{0} }{2} \frac{\widehat A_{2}^{*} }{\widehat A_{2}^{*}\widehat{D}_{W_{2} }^{*} +D_{2}^{2} }
e^{i\phi _{2} } +c.c.=u_{03} +u_{04}
\end{equation}

\begin{equation} \label{eqII16}
 v_{0} =\frac{f_{0} }{2} \frac{\widehat A_{1}^{*} }{\widehat A_{1}^{*}\widehat{D}_{W_{1} }^{*}
+D_{1}^{2} } e^{i\phi _{1} } +c.c.=v_{01} +v_{02}
\end{equation}

\begin{equation} \label{eqII17}
 w_{0} =-\frac{f_{0} }{2} \frac{D_{1} }{\widehat A_{1}^{*}\widehat{D}_{W_{1} }^{*}
+D_{1}^{2} } e^{i\phi _{1} } +\frac{f_{0} }{2} \frac{D_{2} }{\widehat A_{2}^{*}\widehat{D}_{W_{2} }^{*} +D_{2}^{2}
} e^{i\phi _{2} } +c.c.=w_{01} +w_{02} +w_{03} +w_{04}
\end{equation}
where
\begin{equation} \label{eqII18}
\widehat A_{1,2}^{*} =\widehat{D}_{W_{1,2} }^{*} -\frac{\widetilde{Ra} }{\widehat{D}_{\theta_{1,2}
}^{*} } \end{equation}
Components of velocity satisfy the following relations:
\[w_{02} =\left(w_{01}
\right)^{*} , w_{04} =\left(w_{03} \right)^{*} , v_{02} =\left(v_{01} \right)^{*}
, v_{04} =\left(v_{03} \right)^{*} ,  u_{02} =\left(u_{01} \right)^{*} , u_{04} =\left(u_{03} \right)^{*}.  \].

\section*{Appendix III. Calculation of the Reynolds stresses}

To close the equations (\ref{eq15})-(\ref{eq16}) we have to calculate the Reynolds stresses $T^{ik} =\overline{v_{0}^{i} v_{0}^{k} }$ or rather its components:

\begin{equation}\label{eqIII1}
  T^{31} =\overline{w_{0} u_{0} }=\overline{w_{01} \left(u_{01}
\right)^{*} }+\overline{\left(w_{01} \right)^{*} u_{01} }+\overline{w_{03} \left(u_{03}
\right)^{*} }+\overline{\left(w_{03} \right)^{*} u_{03} }
\end{equation}
\begin{equation} \label{eqIII2}
  T^{32} =\overline{w_{0} v_{0} }=\overline{w_{01}
\left(v_{01} \right)^{*} }+\overline{\left(w_{01} \right)^{*} v_{01} }+\overline{w_{03}
\left(v_{03} \right)^{*} }+\overline{\left(w_{03} \right)^{*} v_{03} }
\end{equation}
Substituting the solutions for the small-scale velocity fields (\ref{eqII15})-(\ref{eqII17}) obtained in Appendix II, into the equations (\ref{eqIII1})-(\ref{eqIII2}), we can find the following expression for the correlators:
\begin{equation} \label{eqIII3}
 T^{31} =\frac{f_{0}^{2} }{4} \frac{D_{2}(\widehat A_{2}+\widehat A_{2}^{*})
}{\left|\widehat A_{2}\widehat{D}_{W_{2} } +D_{2}^{2} \right|^{2} }
\end{equation}

\begin{equation}\label{eqIII4}
    T^{32} =-\frac{f_{0}^{2} }{4} \frac{D_{1}(\widehat A_{1}+\widehat A_{1}^{*})  }{\left|\widehat A_{1}\widehat{D}_{W_{1} } +D_{1}^{2} \right|^{2}}
\end{equation}

Then with the definition of the operators (\ref{eqII14}) and (\ref{eqII18}), we write down the series of useful relations for the calculation of  $T^{31} $  and    $T^{32} $ :
\[ \left| {\widehat{D}_{{W}_{1,2}}} \right|^{2}  = \widehat{D}_{{W}_{1,2}} \hat {D}_{{W}_{1,2} }^{*}  = 1 + \widetilde {W}_{1,2}^2,\quad \left| {\widehat{D}_{\theta_{1,2}}} \right|^{2}  = \widehat {D}_{\theta_{1,2} } \widehat {D}_{\theta _{1,2} }^{*}  = \ Pr^{-2}  + \widetilde{W}_{1,2}^{2}, \]

\[\left| {\widehat A_{1,2} } \right|^2  = \hat A_{1,2} \widehat A_{1,2}^*  = 1 + \widetilde W_{1,2}^2  - 2Ra\frac{{1 - \ Pr \widetilde W_{1,2}^2 }}{{1 + \ Pr ^2 \widetilde W_{1,2}^2 }} + \frac{{Ra^2 }}{{1 + \ Pr ^2 \widetilde W_{1,2}^2 }}, \]

\[\widehat D_{W_{1,2} } \widehat D_{\theta _{1,2} }  + \widehat D_{W_{1,2}}^* \widehat D_{\theta _{1,2}}^*  = 2(\ Pr ^{-1}  - \tilde W_{1,2}^2 ),\]

\[\widehat D_{W_{1,2} } \widehat A_{1,2}  + \widehat D_{W_{1,2} }^* \widehat A_{1,2}^*  = 2(1 -\widetilde W_{1,2}^2 ) - 2Ra\frac{{1 + \ Pr \widetilde W_{1,2}^2 }}{{1 + \ Pr^2 \widetilde W_{1,2}^2 }}\]

Here we apply the following designations: $\widetilde{W}_{1} =1-W_{1} $, $\widetilde{W}_{2} =1-W_{2} $ . Using these relations,  we can obtain the following expressions:
\[\widehat A_{1,2}  + \widehat A_{1,2}^*  = 2\left( {1 - \frac{{Ra}}{{1 + \ Pr^2 \widetilde W_{1,2}^2 }}} \right),\]

\begin{equation}\label{eqIII5}\end{equation}
\[\left| {\widehat D_{W_{1,2} } \widehat A_{1,2}  + D_{1,2}^2 } \right|^2  = \left( {1 + \widetilde W_{1,2}^2 } \right)^2  + Ra^2 \frac{{1 + \widetilde W_{1,2}^2 }}{{1 + \ Pr ^2 \widetilde W_{1,2}^2 }} + 2D_{1,2}^2 (1 - \widetilde W_{1,2}^2 ) + \]
 \[ + D_{1,2}^4  - 2Ra\frac{{2 + \widetilde W_{1,2}^2  - \ Pr \widetilde W_{1,2}^4  - (1 + \ Pr \widetilde W_{1,2}^2 )(1 - D_{1,2}^2 )}}{{1 + \ Pr^2 \widetilde W_{1,2}^2 }} \]

Substituting (\ref{eqIII5}) in (\ref{eqIII3})-(\ref{eqIII4}) we can find expressions for the Reynolds stresses in general form. For instance, for the atmosphere the Prandtl number is approximately equal to one  $Pr=1$ . In this case, the expressions for the components of Reynolds stresses are simplified:

\begin{equation}\label{eqIII6}
T^{31}  = \frac{{f_0^2 }}{2}D_2 \frac{{(1 + \widetilde W_2^2  - Ra)}}{{(1 + \widetilde W_2^2 )((1 + \widetilde W_2^2 )^2  + 2(D_2^2  - Ra)(1 - \widetilde W_2^2 ) + (D_2^2  - Ra)^2 )}} \end{equation}

\begin{equation}\label{eqIII7}
T^{32}  = -\frac{{f_0^2 }}{2}D_1 \frac{{(1 + \widetilde W_1^2  - Ra)}}{{(1 + \widetilde W_1^2 )((1 + \widetilde W_1^2 )^2  + 2(D_1^2  - Ra)(1 - \widetilde W_1^2 ) + (D_1^2  - Ra)^2 )}} \end{equation}


\begin{thebibliography}{21}

\bibitem{1s} A.S. Monin. Theoretical fundamentals of geophysical hydrodynamics. Leningrad: Gidrometeoizdat, 1988. 424 p.
\bibitem{2s} O. G. Onishchenko, Pokhotelov O. A.,  Astafieva N. M. Generation of large-scale vortices and zonal winds in atmopheric planets, UFN,178, 605 (2008)

\bibitem{3s} Shmerlin B. Y., Kalashnik M. V. Convective instability of Rayleigh in the presence of phase transitions of the humidity. The formation of the large-scale eddies and cloud structures, UFN, 183, 497 (2013)

\bibitem{4s} V. I. Petviashvili, Pokhotelov O. A. Solitary vortices in plasma and atmosphere. Moscow: Energoatomizdat, 1989. 200 p.

\bibitem{5s} Aburjania G. D. Self-organization of nonlinear vortex structures and vortex turbulence in dispersive media. Moscow: Komkniga, 2006. 328 p.

\bibitem{6s} Kolesnichenko  A.V., Marov  M. Ya., Turbulence and self-organization. Problems of modeling space and natural mediums, Moscow: BINOM, 2009. 648 p.

\bibitem{7s} Anatoli Tur, Vladimir Yanovsky  Coherent Vortex Structures in Fluids and Plasmas. Springer 2017.

\bibitem{8s} Freedman A. M., Khoperskov A.V. Physics of galactic disks. Moscow: FIZMATLIT, 2011. 632 c.

\bibitem{9s} Moiseev S. S., Rutkevitch  P. B., A. V. Tur, V. V. Yanovsky Vortex dynamos in a helical turbulent convection. Sov.Phys.JETP, 67, 294 (1988)

\bibitem{10s} Lypyan E. A., Mazurov A. A., Rutkevitch P. B., Tur A. V. Generation of large-scale vortices through the action of spiral turbulence of a convective nature, Sov.Phys.JETP, 75, 838 (1992)

\bibitem{11s} Moiseev S. S., Oganjan  K. R., Rutkevich P. B., Tur  A.V., Khomenko G. A., Yanovsky V. V. Vortex dynamo in helical turbulence. In the collection: Integrability and kinetic equations for solitons, Sciences.Dumka, Kiev, 1990, pp. 280-382.

\bibitem{12s} Zimin V. D., Levina G. V., Moiseev S. S., Shvarts K. G. Modeling of large-scale vortical processes in a heated from below rotating layer. Dokl. An SSSR, 1990, vol. 312, no. 6, pp. 1372-1374.

\bibitem{13s} G. V. Levina, E. E. Starsev, Zimin V. D., Moiseev S. S., K. G. Bogatirev, Schwarts K. G. Mathematical and laboratory simulation of tropical cyclones. Proc. of the 5-th EPS Liquid State Conference held in Institute for Problem in Mechanics USSR Academy of Sciences. Moscow, October 16-21, 1989, p. 172-175.

\bibitem{14s} A. V. Tur, V. V. Yanovsky.  Large-scale instability in hydrodynamics with stable temperature stratification driven by small-scale helical force.ar Xiv:1204.5024 v.1[physics. Flu-dyn.](2012)

\bibitem{15s} A. V. Tur, V. V. Yanovsky.  Non Linear Vortex Structure in Stratified Driven by Small - scale Helical Forse. Open Journal of Fluid Dynamics, 3, 64-74 (2013)

\bibitem{16s} E. A. Novikov, Functionals and the random force method in turbulence theory. JETP, 1964, V. 47, V. 5(11), pp. 1919-1926

\bibitem{17s} V. I. Klyatskin, Stochastic equations and waves in randomly inhomogeneous media. M.:Nauka, 1980, 337 p.

\bibitem{18s} Frishe U., She Z. S., Sulem P. L. Large Scale Flow Driven by the Anisotropic Kinetic Alpha Effect, Physica D 28, 382 (1987)

\bibitem{19s} G. Rudiger.  On the    $\alpha$ - Effect for Slow and Fast Rotation, Astron. Nachr., 1978, V. 299, No.4, pp. 217-222.

\bibitem{20s} O. G. Chkhetiani. Self-organization and turbulence in a reflection-asymmetric plasma-hydrodynamic environments. Diss. on competition of a scientific degree. academic degree of doctor. Fiz.-Mat. Sciences. Moscow, 1999, 262 p.

\bibitem{21s} M. I. Kopp, A. V. Tur, V. V. Yanovsky. Nonlinear Vortex Structures in Rotating Fluid Obliquely. Open Journal of Fluid Dynamics, 5, 311-321 (2015)

\end{thebibliography}
\end{document}